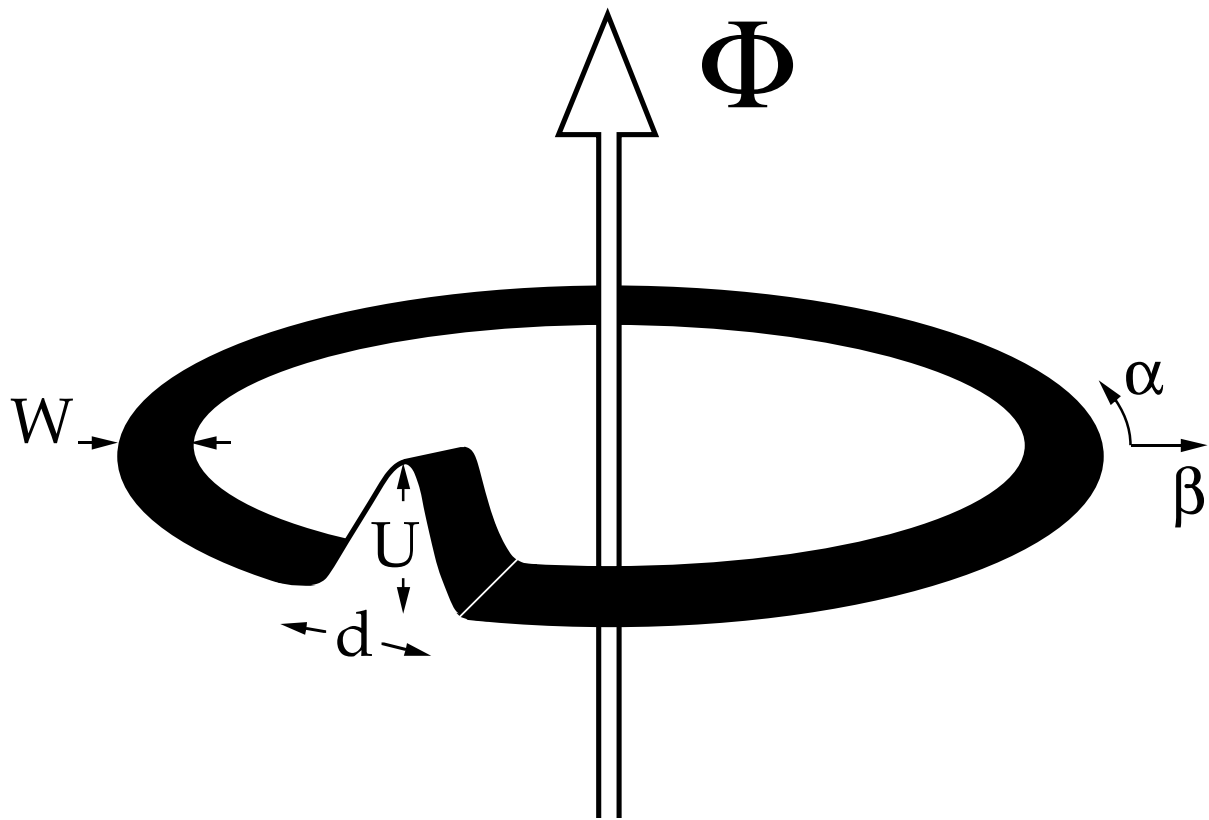

Figure 1

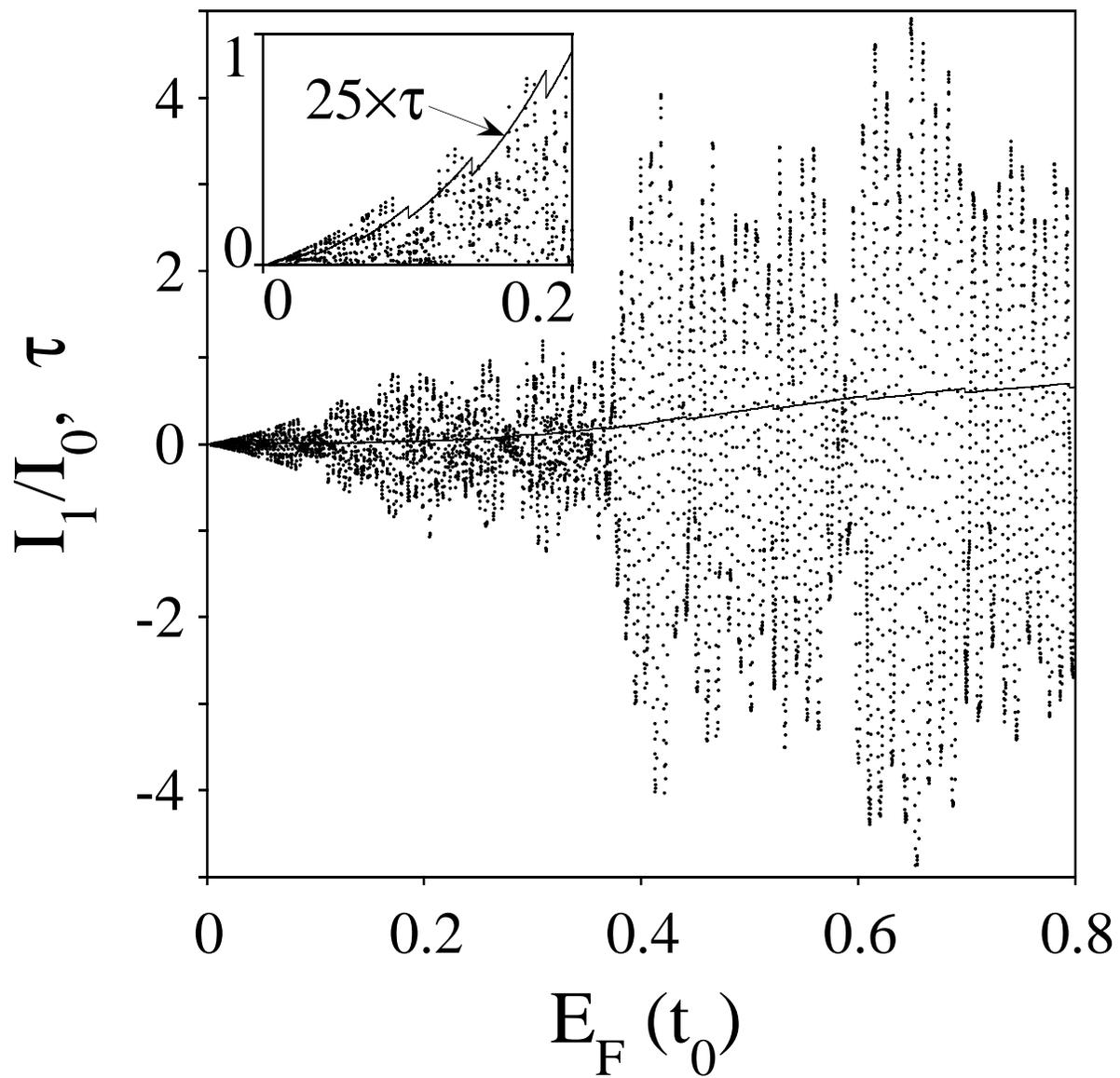

Figure 2

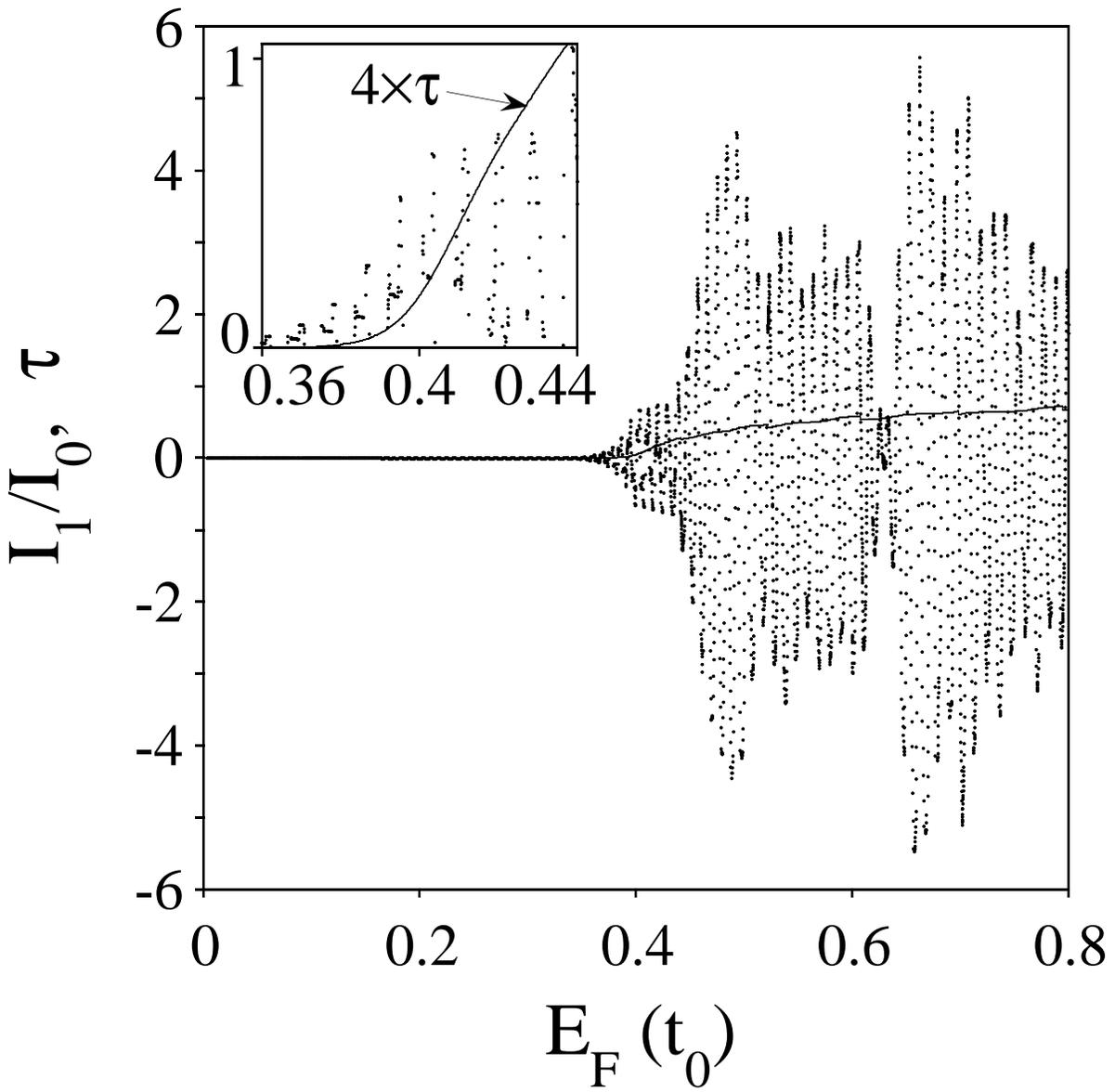

Figure 3

# SEMICONDUCTOR ANALOG OF THE LARGE PERSISTENT CURRENTS OBSERVED IN SMALL GOLD RINGS

George Kirczenow
*Department of Physics, Simon Fraser University,
Burnaby, British Columbia, Canada, V5A 1S6*

The remarkably large persistent currents that are observed in disordered micron-scale gold rings at low temperatures have recently been explained in a theory of non-interacting electrons scattered by crystal grain boundaries. The present article examines the possibility that the basic physics underlying this explanation may also have observable consequences in a different system, a ballistic two-dimensional semiconductor ring with a potential barrier. It is predicted, using computer simulations, that such semiconductor rings can exhibit large persistent currents $\sim ev_F/L$ ($L$ is the ring circumference), despite the electron transmission probability through the barrier being small, in the regime of quantum-mechanical tunneling of electrons through the barrier. This, like the phenomenon observed in gold rings, is a manifestation of the fundamental dissimilarity of non-equilibrium transport phenomena and equilibrium persistent currents.

## 1. Introduction

When a conductor in its ground state is immersed in a magnetic field, it acquires a magnetic moment and an associated circulating electric current. This current is an *equilibrium* property and does not dissipate. It is thus referred to as a "persistent current." Persistent currents were first predicted by London to occur in aromatic molecules [1]. In normal metal rings, they have been discussed theoretically beginning in the 1960's [2]. More recently, Büttiker, Landauer and Imry [3] showed that they should exist even in the presence of disorder. Cheung, Riedel and Gefen [4] estimated the magnitude of the persistent currents that should circulate in mesoscopic normal metal rings in the diffusive regime. They predicted that, at low temperatures, the typical persistent current should be $I_{typ} \sim I_0\, l/L$ where $I_0 = ev_F/L$, $e$ is the electron charge, $L$ is the ring's circumference, $v_F$ is the electron Fermi velocity, and $l$ is the elastic mean free path. Since then, persistent currents have been observed in an array of $10^7$ copper rings by Lévy *et al*., [5] and in individual gold and semiconductor rings by Chandrasekhar et al. [6] and Mailly *et al*. [7] The measured persistent currents in the gold rings [6] were found to be $\sim I_0$, surprisingly large values considering that the mean free paths were quite small, $l/L \sim 0.01$. This experimental finding has stimulated a great deal of interest and debate as to its origin. Much theoretical attention has focussed on the question whether the large observed persistent currents are an effect of electron- electron interactions. But no consensus has been reached as to whether electron-electron interactions can form the basis of a viable explanation, and this issue continues to be highly controversial [8].

Recently, a quite different explanation of the large persistent currents observed in the gold rings has been proposed [9], based on the results of three-dimensional computer simulations and on the known materials properties of thin gold films. -- The previous theoretical work had assumed that the electron scattering in the gold rings was by *random impurities*. However, the rings of Chandrasekhar *et al*. [6] were fabricated from gold films grown on an amorphous substrate (oxidized silicon), and such films are polycrystalline. The low temperature resistance of high quality thin polycrystalline gold films has been found experimentally [10] to be dominated not by random impurity scattering but by *grain boundary* scattering. The recent computer simulations [9] showed that models of rings that incorporate grain boundary scattering differ greatly in their predictions from models based on random impurity scattering. Unlike random impurity models, the grain boundary scattering models were found [9] to explain the experiments of Chandrasekhar *et al*. [6], within a non-interacting electron picture.

It is of interest to know whether the physics underlying the grain-boundary scattering explanation of the gold ring experiments can manifest itself in other systems as well. Since persistent currents have been observed in semiconductor rings by Mailly, Chapelier and Benoit, [7] the possibility that a semiconductor analog of the effect may exist is particularly intriguing, and is explored theoretically in the present article. The semiconductor rings considered here differ from gold rings in that: (i) The semiconductor rings are two-dimensional while the gold rings are three dimensional. (ii) The electron density in the semiconductor rings is much lower and (iii) An electrostatic potential barrier imposed across the ballistic semiconductor ring replaces the crystal grain boundaries of the gold ring.

In Ref. [9], the Landauer picture of electrical resistance [11] was adopted. The resistance of the ring was expressed in terms of the total multi-channel transmission $T = \text{Tr}(tt^\dagger)$ around the ring of electrons at the Fermi energy. [*For the purpose of defining T*, the ring can be imagined to be severed and the ends thus created to be connected to ideal leads that inject and absorb electrons.] If $\eta$ is the number of conducting channels (transverse modes) in the ring at the Fermi energy, $\tau \equiv T/\eta$ is the probability that a Fermi energy electron is transmitted all the way around the ring. The relevant experimental results of Chandrasekhar *et al*. [6] can then be summarized as follows [9]: Gold rings with small values of $\tau \sim 0.01$ were observed to display large persistent currents, of order $I_0$. The question to be considered here is whether a ballistic semiconductor ring with an electrostatic potential barrier can display similarly large persistent currents for low electron transmission probabilities $\tau$ through the barrier.

## 2. Model and method of solution

The model studied here is a lattice Hamiltonian $H$ describing electrons in a narrow two-dimensional ring with a potential barrier, and threaded by a magnetic flux:

$$H = \sum_{j,k} u_{jk} a^\dagger_{jk} a_{jk} - \zeta t^\alpha_{jk} a^\dagger_{jk} a_{j+1\,k} - t^\beta_{jk} a^\dagger_{jk} a_{j\,k+1}$$

+ Hermitian conjugate

Here $\zeta = \exp(2\pi i \Phi / Z)$, $\Phi$ is the magnetic flux threading the ring in units of the flux quantum $h/e$. $Z$ is the circumference of the ring in units of the lattice parameter. $a^\dagger_{jk}$ is the electron creation operator at site $jk$. The electron spin index is suppressed and the Zeeman splitting between spin up and down is neglected. The first term in $H$ is the electron on-site energy. The second term describes electron hopping in the azimuthal direction around the ring; the third term describes hopping in the orthogonal direction. Periodic boundary conditions in the $\alpha$ (or $j$) direction close the linear strip to form a ring. (See Fig.1, for the hopping directions $\alpha$ and $\beta$ in the ring geometry.) Hamiltonians of this general form have been used previously in numerical studies of persistent currents in narrow rings with random defect scattering [12]. In Ref. [9], an analogous three-dimensional Hamiltonian was used to study the effects of scattering by grain boundaries in metal rings. The grain boundaries in the ring were modelled by setting the hopping coefficients $t$ that represent hopping across a grain boundary smaller than the other hopping coefficients which were all chosen to be equal. For the ballistic semiconductor rings to be considered here, all of the hopping coefficients are taken to be equal,

$$t^\alpha_{jk} = t^\beta_{jk} = t_0,$$

and the electrostatic potential barrier is represented by an appropriate choice of the electron on-site energies $u_{jk}$ (see below).

The persistent current $I$ was evaluated at zero temperature by finding the eigenvalues of $H$ numerically, and using the well known result [2] $I = -e/h\, \partial E/\partial \Phi$, where $E$ is the total electronic ground state energy of the ring. Since the persistent current is periodic in the flux $\Phi$,

$$I = \sum_{n=1}^{\infty} I_n \sin(2\pi n \Phi),$$

the first Fourier coefficient $I_1$ which is characteristic of the amplitude of the persistent current, will be considered here. This Fourier coefficient was evaluated assuming that the number of electrons in the ring is independent of the value of the magnetic flux, and that the ring is kept in its ground state. The barrier transmission probabilities $\tau$ were calculated numerically at zero magnetic flux by solving the electron Lippmann-Schwinger equation numerically [13].

## 3. Results

The numerical results presented here are for rings of width $W = 50$ sites (see Fig.1), and circumference $Z = 800$ sites. The smooth functional form

$$u_{jk} = \begin{cases} \frac{U}{2}\left[1 + \cos\left(2\pi\left(\frac{j-1}{d-1} - \frac{1}{2}\right)\right)\right] & 1 \leq j \leq d \\ 0 & \text{otherwise} \end{cases}$$

was used to model the electrostatic potential energy barrier. Here $U$ is the barrier's height and $d$ is its total thickness (as depicted in Fig.1).

Figures 2 and 3 show the results of the calculations of the first Fourier coefficient $I_1$ of the persistent current ($I_1/I_0$ is shown by the dots [14]), as a function of the Fermi energy $E_F$. The solid curve is the electron transmission probability $\tau$ of the barrier at the Fermi energy. The abrupt jumps in $\tau$ occur where the Fermi energy crosses the bottom of a subband and $\eta$ changes. In the figures, $E_F$ is defined to be the energy of the highest occupied single-particle eigenstate of the Hamiltonian $H$ at $\Phi = 0$, and is measured from the bottom of the conduction band. Note that in this model the total energy width of the 2D conduction band is $8t_0$. In both figures the height of the barrier is $U = 0.4t_0$. In Fig. 2 the barrier is narrow with a total thickness $d = 10$ sites; in Fig.3 it is wide with $d = 100$. Each dot in the figures indicates the value of $I_1/I_0$ for a particular number of electrons in the ring. The number of electrons in the ring at $E_F = 0.8t_0$ is 5222 in Fig.2 and 5070 in Fig.3. At $E_F = 0.4t_0$, when the Fermi energy crosses the top of the potential barrier, 10 transverse subbands (each containing spin up and down electrons) are populated *far from* the barrier; for $E_F = 0.8t_0$, 15 transverse subbands are populated.

The persistent current oscillates as the number of electrons in the ring changes, as has been found in previous work [1]-[4], [8], [9], [12]. The amplitude of the oscillations of $I_1/I_0$ is significantly larger than $\tau$ everywhere in the figures. However $\tau$ is *drastically* smaller than $I_1/I_0$, approaching the behavior found in the gold rings [6], [9], only in (or very near to) the quantum tunneling regime where the Fermi level is below the top of the potential barrier, i.e., for $E_F \leq U = 0.4t_0$. This is illustrated in the insets of Figures 2 and 3.

For the thinner tunnel barrier (inset of Fig.2) strong persistent currents ($I_1 \sim I_0$) are seen together with very small transmission probabilities $\tau \sim 0.04$ for $E_F \sim 0.2t_0$. Further into the tunneling regime, at still lower $E_F$, $I_1/I_0$ also becomes small but the amplitude of its oscillations continues to exceed $\tau$ by a factor of 25 - 50.

For the thicker tunnel barrier (inset of Fig.3) strong persistent currents ($I_1 \sim 0.67 I_0$) are seen together with weak transmission probabilities $\tau \sim 0.04$ near the crossover into the tunneling regime at $E_F = 0.4t_0$. As the Fermi energy decreases further and the system moves deeper into the tunneling regime, the transmission probability $\tau$ decreases much more rapidly than the amplitude of the persistent current, although the latter also quickly becomes very small. This is clearly visible in the inset of Fig.3, or one can compare the above numbers for $E_F \sim 0.4t_0$ with those at $E_F \sim 0.36t_0$, where $\tau \sim 0.008 I_1/I_0$ and $I_1/I_0 \sim 0.016$.

## 4. Discussion

The computer simulations reported in Ref. [9] showed clearly that the fundamental reason why grain boundary scattering models are able to explain the gold ring experiments of Chandrasekhar *et al*. [6], while random defect models cannot, is that equilibrium persistent currents and non-equilibrium electron transport are very different phenomena. In Ref. [9] this difference appeared as different dependences of the electron transmission probability $\tau$ (a transport property) and of the persistent current, on the arrangement of defects in the ring. Assembling random defects into a coherent grain boundary crossing the ring was found to depress $\tau$ much more than $I_1/I_0$, with the result that, in a ring with grain boundaries, large persistent currents could coexist with small transport mean free paths [9], as was observed in the experiments [6].

The present study once again illustrates the fundamental dissimilarity of persistent currents and non-equilibrium transport, but this time in the regime of quantum tunneling through an electrostatic tunnel barrier. The tunnel barrier depresses the transmission probability $\tau$ much more than it does the persistent current, making it possible, in principle, to observe large persistent currents $\sim I_0$ coexisting with small barrier transmission probabilities $\tau$. This is the semiconductor analog of the physics observed by Chandrasekhar *et al*. [6] in their gold ring experiments. Furthermore, by varying the strength of the tunnel barrier, it should be possible to observe the more rapid suppression of $\tau$ than $I_1/I_0$ as one progresses deeper into the tunneling regime, as has been predicted here. Such experiments would be of considerable interest. It should be emphasized that the most interesting regime for measurements is that of quantum tunneling through the barrier, rather than classical transmission over it.

The experiments proposed above should be feasible in a device similar to the rings of Mailly, Chapelier and Benoit, [7], but with an additional gate added to set up the potential barrier and to vary it independently of the ring geometry. Other independent gates would also be needed to switch the device from the persistent current (ring) geometry to a quantum wire geometry in which the transmission of the barrier can be measured directly. The technology necessary for making devices with the independently controllable submicron scale gates that are needed for such experiments already exists [15].

*Acknowledgement* — I wish to thank B. L. Johnson and D. Loss for interesting discussions. This work was supported by the Natural Sciences and Engineering Research Council of Canada.

as the subject of the message. Unpacking instructions are at the beginning of the file.

### FIGURE CAPTIONS

**Fig.1** Schematic drawing of a two-dimensional semiconductor ring of width $W$ threaded by a magnetic flux $\Phi$, and with a potential energy barrier of height $U$ and thickness $d$.

**Fig.2** Normalized first Fourier coefficient of the persistent current $I_1/I_0$ (the dots) in a two-dimensional semiconductor ring with a thin potential barrier, and probability $\tau$ of transmission of a Fermi energy electron through the barrier (solid line), vs. Fermi energy $E_F$. The ring has a circumference $Z = 800$ lattice sites and width $W = 50$ sites. The thickness $d$ of the smooth barrier is 10 and its height is $U=0.4t_0$. $E_F$ is in units of $t_0$. Inset: Detail of the plot for small $E_F$.

**Fig.3** Normalized first Fourier coefficient of the persistent current $I_1/I_0$ (the dots) in a two-dimensional semiconductor ring with a thicker potential barrier, and probability $\tau$ of transmission of a Fermi energy electron through the barrier (solid line), vs. Fermi energy $E_F$. Dimensions of the ring are as in Fig. 2. The thickness $d$ of the barrier is 100 sites and its height is $U=0.4t_0$. $E_F$ is in units of $t_0$. Inset: Detail of the plot near where the Fermi energy crosses the top of the barrier at $E_F = 0.4t_0$.